\documentstyle[12pt,a4,a4wide,epsfig]{article}
\def\Journal#1#2#3#4{{#1} {\bf #2}, #3 (#4)}

\def\NPB{{\em Nucl. Phys.} {\bf B}}

\def\PRL{\em Phys. Rev. Lett.}

\def\PRP{{\em Phys. Rep. }}
\def\EPC{{\em Eur. Phys. J.} {\bf C}}

\def\ra{\rightarrow}
\def\be{\begin{equation}}
\def\ee{\end{equation}}

\newcommand{\ls}{\mbox{$\stackrel{<}{\sim}$ }}

\newcommand{\onbb}{neutrinoless double beta decay }
\newcommand{\majo}{Majorana }

\newcommand{\ssm}{see--saw--mechanism }

\newcommand{\nel}{\mbox{$\nu_e$} }
\newcommand{\nmu}{\mbox{$\nu_\mu$} }
\newcommand{\ntau}{\mbox{$\nu_\tau$} }

\newcommand{\Slash}[1]{\mbox{$#1\hspace{-.6em}/$}}
\newcommand{\ba}{\begin{array}{c}}
\newcommand{\baz}{\begin{array}{cc}}
\newcommand{\bad}{\begin{array}{ccc}}
\newcommand{\bea}{\begin{equation} \begin{array}{c}}
\newcommand{\eea}{ \end{array} \end{equation}}
\newcommand{\ea}{\end{array}}

\newcommand{\aen}{\mbox{$\overline{\nu_e} $}}

\newcommand{\nnu}{\mbox{$0\nu \beta \beta$ }}

\newcommand{\mab}{\mbox{$\langle m_{\alpha \beta} \rangle $ }}
\newcommand{\mmm}{\mbox{$\langle m_{\mu \mu} \rangle $ }}
\newcommand{\mtta}{\mbox{$\langle m_{\tau \tau} \rangle $ }}
\newcommand{\mmt}{\mbox{$\langle m_{\mu \tau} \rangle $ }}
\newcommand{\mee}{\mbox{$\langle m_{ee} \rangle $ }}
\newcommand{\met}{\mbox{$\langle m_{e \tau} \rangle $ }}
\newcommand{\meu}{\mbox{$\langle m_{e \mu} \rangle $ }}
\begin{document}
\newpage
\title{\hfill { \bf {\normalsize DO--TH 99/20}}\\ 
\hfill { \bf {\normalsize hep-ph/9911298}}\\ \vskip 1.5cm
\bf Bounds on effective Majorana neutrino masses at HERA}
\author{M. Flanz$^a$
\footnote{Email address: flanz@dilbert.physik.uni-dortmund.de}, W. Rodejohann$^a$
\footnote{Email address: rodejoha@dilbert.physik.uni-dortmund.de}, K. Zuber$^b$
\footnote{Email address: zuber@physik.uni-dortmund.de}\\
{\it \normalsize $^a$Lehrstuhl f\"ur Theoretische Physik III,}\\ 
{\it \normalsize $^b$Lehrstuhl f\"ur Experimentelle Physik IV,}\\
{\it \normalsize Universit\"at Dortmund, Otto--Hahn Str. 4,}\\ 
{\it \normalsize 44221 Dortmund, Germany}}
\date{}
\maketitle
\begin{abstract}
The lepton--number violating process 
$e^{\pm} p \rightarrow \nu_e (\overline{\nu_e}) 
l^{\pm} l'^{\pm} X$ 
mediated by Majorana neutrinos is studied for the HERA collider 
for $ (l l') = (e \tau) , \; (\mu  \tau) , \;(\mu \mu)$ 
and $(\tau \tau)$. Only the muonic decay of the $\tau$ is considered. 
The direct limit on the effective 
muon Majorana mass, $\langle m_{\mu \mu} \rangle$ is 
improved significantly to $4.0 \cdot 10^{3}$ 
GeV and for the first time direct limits on the analogous 
effective masses connected with the tau sector are given, 
namely $4.2 \cdot 10^{3}$ GeV for 
$\langle m_{e \tau} \rangle$,  $4.4 \cdot 10^{3}$ GeV 
for $\langle m_{\mu \tau} \rangle$ and 
$2.0 \cdot 10^{4}$ GeV for $\langle m_{\tau \tau} \rangle$. 
We find that a more general analysis for an upgraded HERA could 
improve this values by a factor of up to 40, yet still being 
orders of magnitude worse than indirect limits.    
\end{abstract}
{\small lepton--hadron processes; massive neutrinos; mass bounds; Majorana neutrinos}

\newpage
\section{Introduction}
\begin{figure}[ht]
\setlength{\unitlength}{1cm}
\begin{center}
\epsfig{file=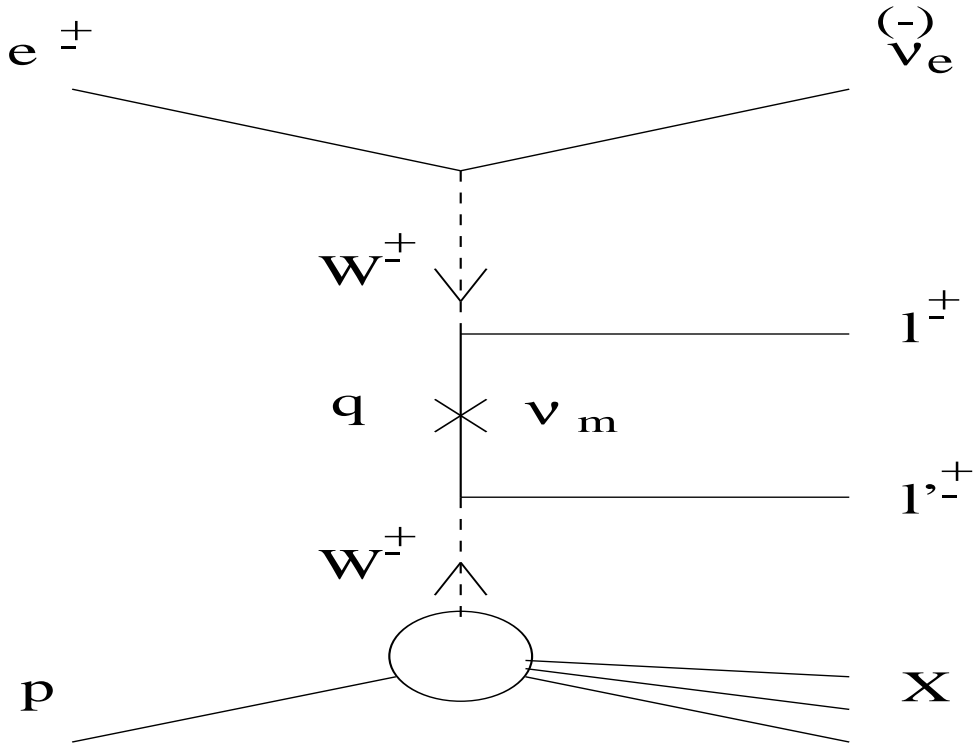,width=13cm,height=10cm}
\end{center}
\caption{\label{feydia}Feynman diagram for the process 
$e^{\pm} p \ra \stackrel{(-)}{\nel} l^{\pm} l'^{\pm} X$. $q$ 
denotes the four--momentum of the propagating Majorana neutrino. Note that 
there is a crossed term and for $l \neq l'$ two possible assignments for 
the corresponding lepton vertex exist.}
\end{figure}

Since there is growing experimental evidence \cite{kaireport} 
of nonzero neutrino masses using neutrino oscillations as explanations 
\cite{exprev}, 
an additional fundamental question still to be solved
is the character of
the neutrinos, i.\ e.\ 
are they Dirac or Majorana particles? 
From the theoretical side the latter case is favored since 
Majorana particles pop out of almost every GUT \cite{GUT} and are 
also the product of the attractive \ssm \cite{seesaw}. 
The most important tool to answer this question is the 
detection of lepton--number violation in the neutrino sector. 
The most effort of theoretical and experimental 
work has been put in \onbb ($\nnu \!$), resulting in an upper limit on the 
effective \majo{}mass  
$\mee \! \! = | \sum U_{em}^2 m_m \eta^{\rm CP}_m |$ 
of about 0.2 eV \cite{hdmo},
where $m_m$ are the mass eigenvalues, $\eta^{\rm CP}_m = \pm 1$ 
the relative CP parities and $U_{em}$ the mixing matrix
elements. In general, there is a $3 \times 3$ matrix of effective 
Majorana masses, the elements being 
\be \label{meffmatrix}
\mab  = 
\left| \sum U_{\alpha m} U_{\beta m} m_m \eta^{\rm CP}_m  \right|
\mbox{ with } 
\alpha, \, \beta = e , \, \mu , \, \tau  .   
\ee  
For sake of simplicity we assume that the elements $U_{\alpha m}$ 
are real and skip also $\eta^{\rm CP}_m$.  
Only few direct information on elements other than $\mee \! \!$ is available: 
muon--positron conversion in sulfur gives a limit on 
$\langle m_{\mu e} \rangle \ls 0.4$ (1.9) GeV, when the final state 
proton pairs are in a spin singlet (triplet) state, respectively. This 
limit is obtained when comparing the theoretical value from \cite{doi} 
with the PDG branching ratio limit \cite{PDG} and using the fact that the 
matrix element of the process is proportional to 
$|\langle m_{\mu e} \rangle|^2$. 
In a recent paper \cite{FRZ} we considered the reaction 
$\nmu N \ra \mu^- \mu^+ \mu^+ X $, mediated by Majorana neutrinos, 
and deduced a limit of $\mmm \ls 10^4$ GeV, improving 
the previous bound \cite{japaner} by one order of magnitude. 
To our knowledge, there are no direct limits on other elements of 
$\mab \! \!$. 
Note that we are considering direct limits, i.\ e.\ using 
processes sensitive on the respective quantity. Indirect bounds, 
obtained from oscillation experiments and unitarity of the mixing matrix 
will of course be far more restrictive.\\ 
In this paper we will study the process  
\be  \label{process}
e^{\pm} p \ra \stackrel{(-)}{\nel}
l^{\pm} l'^{\pm} X, \mbox{ with } (l l') = (e \tau) , \; 
 (\mu  \tau) , \;(\mu  \mu) \mbox{ and } (\tau \tau) 
\ee
for the case of the HERA collider. We will focus mainly 
on the $e^+ p$ mode, but the qualitative conclusions 
we draw remain of course valid for $e^-p$--collisions as well. 
We demand the taus\footnote{We shall use the term electron, muon or 
tau for both, particle and antiparticle.} to decay in 
muons to take advantage of the like--sign lepton signature, which is more 
unique for muons than electrons. A further effect is
to have a small number of
like--sign ($e \mu$) or ($ee$) events, which might be 
more background dominated.  
The relevant diagram for process (\ref{process}) is 
shown in Fig.\ \ref{feydia}. 

It is evident that such a process has a spectacular signature with 
large missing transverse momentum ($\Slash{p}_T$) and two like--sign 
leptons, isolated from the hadronic remnants. 
Direct production of heavy Majoranas $N$ at HERA has been 
studied before \cite{buchgreub} with the process 
$e^- p \ra X N \ra X W^{\pm} l^{\mp} \ra X \nu l'^{\pm} l^{\mp}$ 
resulting in two leptons with different charge. 
In contrast to direct production the process discussed in the 
present paper deserves some attention because of its unique
signature.

\section{Analysis}
We use HERA kinematics for 
electron and proton energies 
$E_e$ = 27.5 GeV, $E_p$ = 820 GeV and the GRV 98 \cite{grv} set of parton 
distributions. 
To mimic the experimental situation the following 
kinematical cuts are applied: 
$| \eta_{l,  X} | \le 2.0$ and $| \Slash{p}_T | \ge 10$ GeV\@. 
Here $| \Slash{p}_T | = \sqrt{ (\sum p_x^{\nu})^2 + (\sum p_y^{\nu})^2}$ is 
the total missing transverse momentum, 
with the sum going over all neutrinos, 
i.\ e.\ one neutrino for the $(\mu \mu)$, three for the 
$(e \tau)$ and $(\mu \tau)$ and five for the $(\tau \tau)$ case.      
$\eta_{l,X} = - \ln \tan (\theta_{l ,X}/2)$  
is the pseudorapidity of the charged lepton $l$ and 
the hadronic final state $X$ respectively, with $\theta$ the polar
angle 
in a system where the $z$--axis is parallel to the proton direction. 
In addition, we want the charged leptons to be 
isolated from the hadrons and demand 
$\Delta R = \sqrt{(\phi_l - \phi_X)^2 + (\eta_l - \eta_X)^2} > 0.5$ with 
$\phi$ being the azimuthal angle.\\
The exact calculation of the diagram and some general features of 
the resulting cross section are described in \cite{FRZ}. 
For the problem at hand we additionally 
folded in the three--body decay of the tau leptons.
When considering heavy neutrinos, one has to note the mixing of the 
usual standard model (SM) leptons with these hypothetical particles. 
The Lagrangian for the lepton $l$ coupling to neutrino 
mass eigenstates $\nu_m$ is:   
\be
\mbox{${\cal L}$} = \frac{g}{\sqrt{2}} \sum_m U_{l m} \overline{l} 
\gamma_{\alpha} \gamma_- \nu_m W^{\alpha}  \; \; + \; \; {\rm h. \; c.}  
\ee
This leads to a dependence of the cross sections of the form: 
\be
d \sigma ( ep \rightarrow \nu_e l l' X) 
\propto \left| \sum U_{lm} U_{l' m} \frac{m_m}{q^2 - m_m^2} \right|^2
\ee
with $q$ being the four--momentum of the propagating Majorana neutrino.\\ 
The DELPHI collaboration \cite{delphi} 
examined the mode $Z \ra \overline{\nu_l} \nu_m$ and found a limit of 
$|U_{l m}|^2 < 2 \cdot 10^{-5}$ for masses up to $m_m \simeq 80$ GeV and 
$l = e, \mu$ and $\tau$.  
For larger masses analyses of neutrino--nucleon scattering
and other processes yielded \cite{Ulimits} 
\be \label{ulim}
\sum |U_{e m}|^2 < 6.6 \cdot 10^{-3} , \; 
\sum|U_{\mu m}|^2 < 6.0 \cdot 10^{-3}   \mbox{ and } 
\sum |U_{\tau m}|^2 < 1.8 \cdot 10^{-2} .
\ee

\section{Results and Discussion}
\begin{figure}[ht]
\setlength{\unitlength}{1cm}
\begin{center}
\epsfig{file=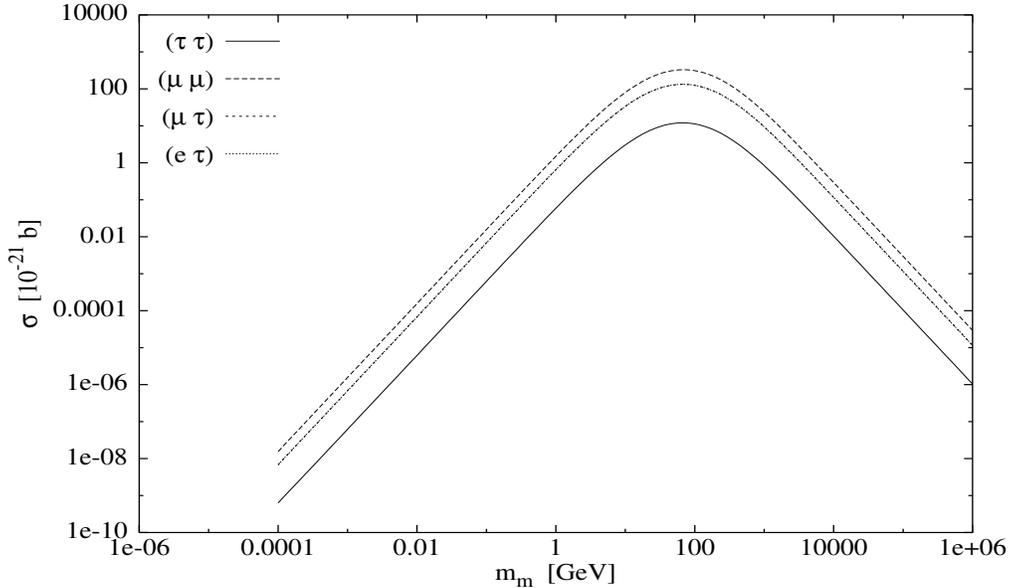,width=13cm,height=8cm}
\end{center}
\vspace{-0.5cm}
\caption{\label{sigvonm}Total cross section for 
$e^{+} p \ra \aen l^{+} l'^{+} X$ as a function of {\it one} eigenvalue 
$m_m$. No limits on $U_{l m}$ are applied, the branching ratio 
for taus into muons is included. The $(e \tau)$ and 
 $(\mu \tau)$ cases are indistinguishable in this plot.}
\end{figure}

In Fig.\ \ref{sigvonm} the total cross section as a function of 
{\it one} mass eigenvalue $m_m$ is shown for all combinations of 
final state charged leptons, 
without considering the above mentioned $U_{l m}$ limits 
(i.\ e.\ setting $|U_{l m}|^2 = 1$). Our condition that the 
tau decays into a muon is included. 
As can be seen, there is a maximum at about 70 GeV, which has 
purely kinematical reasons, see \cite{FRZ}. This means that we can assume 
one eigenvalue dominating the sum $\sum m_m^2 (q^2 - m_m^2)^{-2}$. 
For small masses the cross section rises with 
$m_m^2$ and for higher masses it 
falls with $m_m^{-2}$ as can be 
understood from the two extreme limits of 
$m_m^2 (q^2 - m_m^2)^{-2}$. 
The masses of the final state leptons have almost no effect 
(less than 5 $\%$), so that 
the only numerical difference comes from the branching ratios (BR),   
the $U_{l m}$ limits from Eq.\ (\ref{ulim}) and the factor 2 for 
the $(e \tau)$ and $(\mu \tau)$ cases. The latter comes from the 
two possible assignments the two leptons have, when emitted from the 
Majorana vertex. 

\begin{figure}[hp]
\setlength{\unitlength}{1cm}
\vspace{-2.2cm}
\begin{center}
\epsfig{file=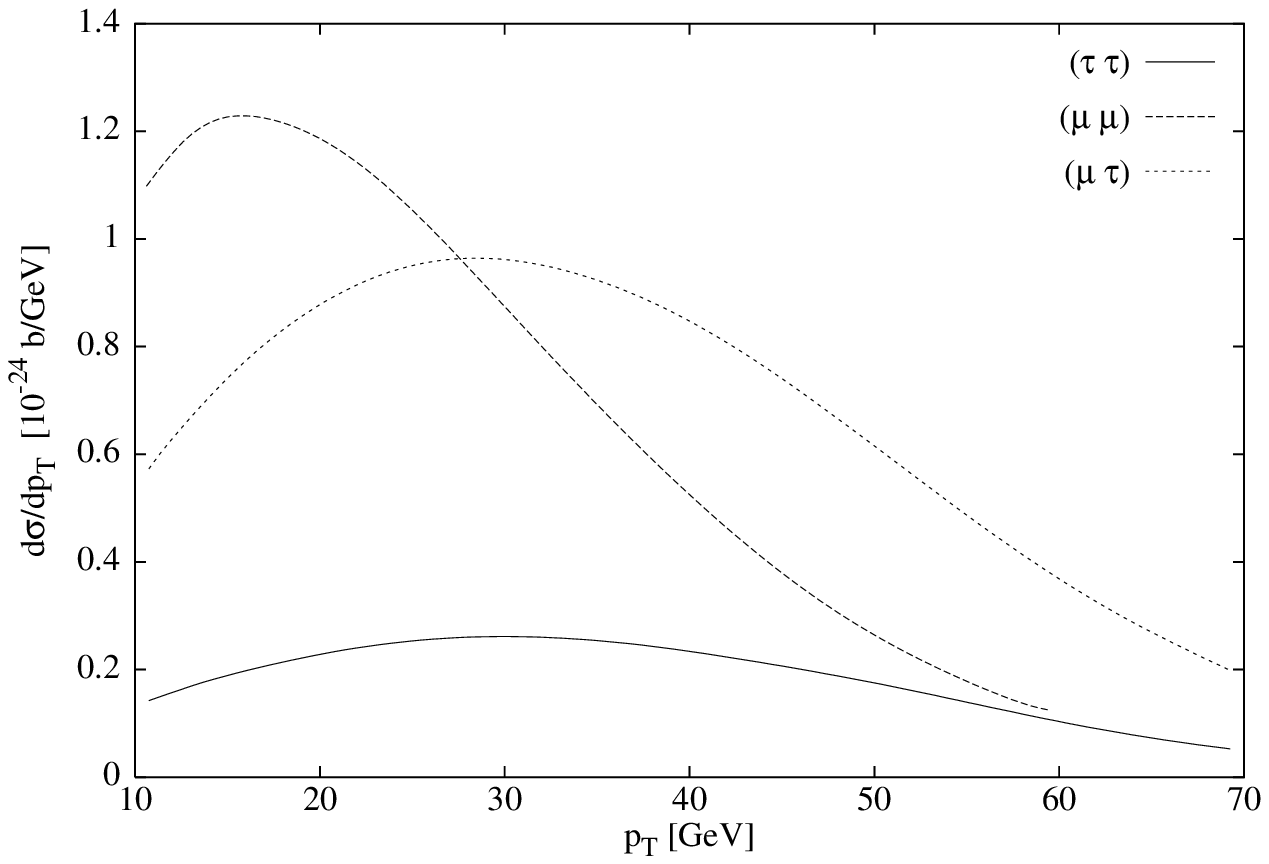,width=13cm,height=8cm}
\end{center}
\vspace{-0.5cm}
\caption{\label{dsigdpt}Distribution of the total missing 
transverse momentum $\Slash{p}_T$ for the 
$(\mu \tau)$, $(\mu \mu)$ and $(\tau \tau)$ case in the reaction 
$e^{+} p \ra \aen l^{+} l'^{+} X$ for $m_m = 80$ 
GeV\@. In order to have the curves in the same order of magnitude we did not 
include the tau branching ratio.} 
\vspace{0.5cm}
\setlength{\unitlength}{1cm}
\begin{center}
\epsfig{file=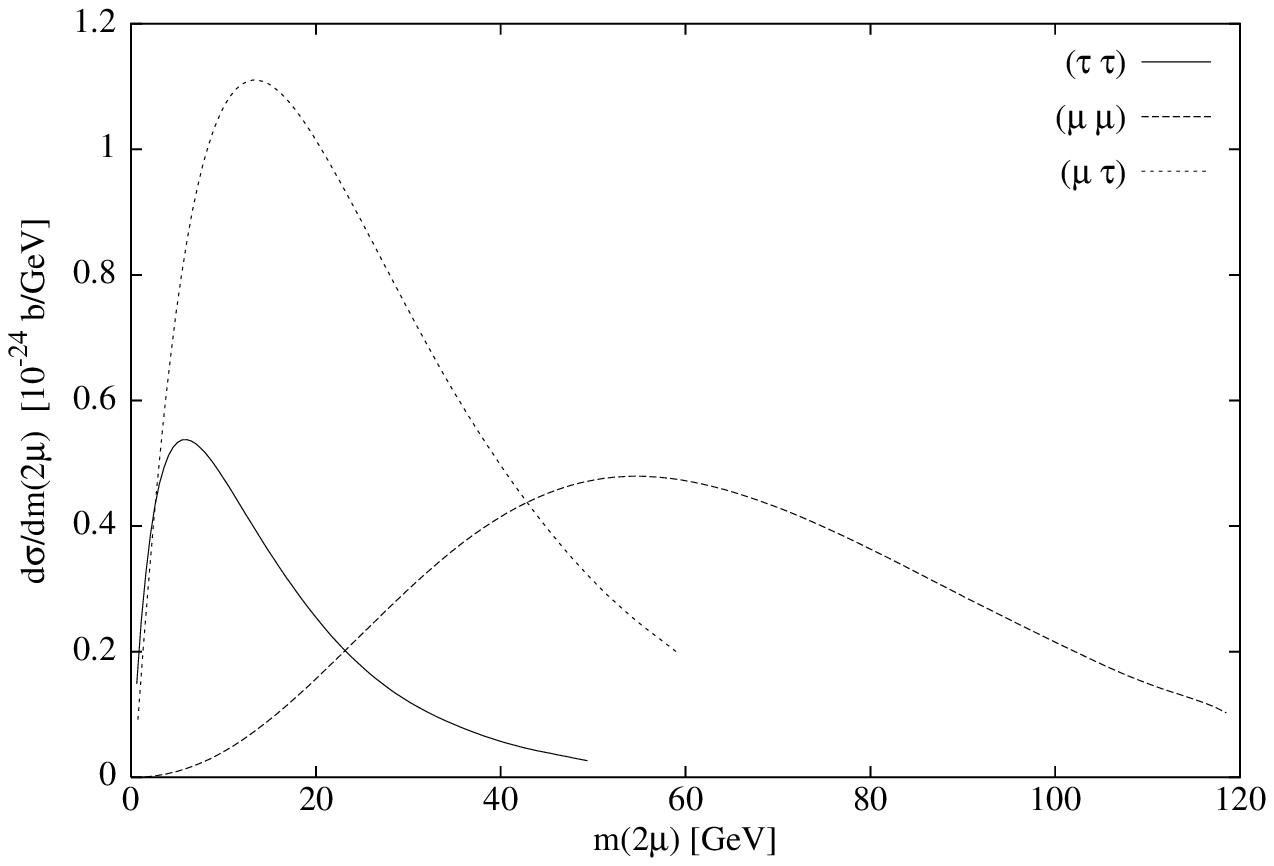,width=13cm,height=8cm}
\end{center}
\vspace{-0.7cm}
\caption{\label{dsigdminv}Invariant mass distribution for the two 
like--sign muons given for the same parameters as in the previous figure.}
\end{figure}

These limits combined with the branching ratio 
(BR($\tau \ra \mu \nmu \ntau$) = 0.1732 \cite{PDG}) 
lead to a maximal cross section of about $10^{-23}$ b in the  
$(\mu \mu)$ case for a neutrino mass of 80 GeV\@. 
It is about 12 orders of magnitude smaller than the SM charged current
(CC) $e^+ p$ cross section of 
$\sigma_{\rm CC}(e^+ p , \; Q^2 > 200 \, {\rm GeV}^2) \simeq  30$ 
pb \cite{Zeus} (we checked that the $Q^2$ condition is not 
significantly violated for the cuts we applied). 
Nevertheless the above cross section is some orders of magnitude 
closer to the relevant SM CC process than most other exotic 
Majorana neutrino induced $\Delta L = 2$ processes 
such as $K^+ \ra \pi^- \mu^+ \mu^+$ \cite{Kpimumu}
or $\mu^-  \mu^+$--conversion via muon--capture in $^{44}$Ti \cite{Moha}. 
These have ratios with respect to the relevant SM CC process of at most 
$10^{-20}$. For $\nmu N \ra \mu^- \mu^+ \mu^+ X $   
a ratio of $10^{-17}$ \cite{FRZ} for a 500 GeV neutrino beam
is achieved, so the process described here results in an 
improvement of another 5 orders of magnitude.\\ 
The $e^- p$ mode gives (in contrast to the normal CC process) 
a cross section {\it smaller} 
than for $e^+ p$ by a factor of about 2 (1.9 for small masses, 
2.5 for masses higher than $10^2$ GeV), so that the ratio for this mode 
is a factor 4 worse. The other cases, like $(\tau \tau$), have ratios 
with respect to the SM CC process maximally one order of magnitude 
smaller than the ($\mu \mu$) case.\\ 
As an example for differential cross sections we plot in 
Fig.\ \ref{dsigdpt} for $m_m = 80$ GeV the 
distribution of the missing transverse momentum for the 
$(\mu \mu)$, $(\mu \tau)$ and $(\tau \tau)$ case. 
Note that all these cases have two like--sign muons in the final state. 
The mean values are  
$\langle \Slash{p}_T \rangle \simeq 28.3 , \; 37.0 \mbox{ and } 37.3$ GeV, 
respectively. 
The shape is different for each case, 
despite the similar mean value of missing transverse momentum. 
To distinguish, say, $(\mu \tau)$ from $(\tau
\tau)$ events, one should consider other distributions, e.\ g.\  
the invariant mass of the two muons, $m (2 \mu)$, 
as displayed in Fig.\ \ref{dsigdminv}. $\!$Whether a muon comes 
directly from the Majorana vertex or from the tau--decay makes its 
energy and momenta fraction of the total available energy smaller and 
changes its invariant mass. Here the mean values are 
$\langle m (2 \mu) \rangle \simeq 66.8 , \; 28.1 \mbox{ and } 16.1$ GeV,
for $(\mu \mu)$, $(\mu \tau)$ and $(\tau \tau)$, respectively.
Unfortunately, this procedure requires high statistics. On an 
event--by--event analysis detailed kinematic reconstruction as well as 
angular distributions might be more useful.
\newpage

Because of its lepton--number violating character 
the discussed process should be background free from 
standard model processes. On the other hand, misidentification
of the muon charge might create some like--sign dimuon events. 
Processes producing opposite sign muon pairs are pair 
production of heavy quarks, photon--photon interactions,
$Z$--boson production, Drell--Yan pairs from resolved 
photon--proton interactions, pion punchthrough
associated with a single muon event and beam related background. 
Practically all these processes
can be eliminated by kinematical arguments. 
Therefore the identification of two like--sign, isolated
muons with large $p_T$ in addition to a large missing 
transverse momentum should indeed be an outstanding
signature. 
For the ($e \tau$) channel, which has a like--sign 
$e \mu$ signature, single $W$ production \cite{Wprod} is 
a severe background. However, sophisticated 
kinematical arguments as those given in \cite{buchgreub} for the 
case of direct Majorana production 
might be also applicable in the case discussed here. 
Because of this we shall assume zero background, leaving a more 
detailed analysis for further studies. In case of observation 
such a detailed analysis has to be done anyway in order to rule 
out any standard model process as the ones described above.\\  
In order to get bounds for \mab we assume that the cross sections 
displayed in Fig.\ \ref{sigvonm}  
are proportional to $|\mab \! \!|^2$
and take the luminosities used in 
searches for isolated lepton events with missing transverse 
momentum, i.\ e.\ 
$\mbox{${\cal L}$}_{e^+} = 36.5  $ pb$^{-1}$ (H1, \cite{isollepH1}) and 
$\mbox{${\cal L}$}_{e^+} = 47.7  $ pb$^{-1}$ (ZEUS, \cite{isollepZEUS}). 
We take the average of 42.1 pb$^{-1}$ and get values in the 
range of $10^{3}$ to $10^{4}$ GeV, thereby 
improving the \mmm limit with respect to \cite{FRZ} significantly  
and giving for the first time direct 
limits on $\langle m_{\alpha \tau} \rangle$. 
Combining all limits, ignoring 
possible phases in the elements   
$U_{\alpha m}$ (therefore getting a symmetrical matrix $\mab \! \!$) 
as well as skipping the intrinsic CP parities,
the following direct bounds for the effective 
Majorana mass matrix exist: 
\be \label{meffresult}
\mab = \left( \bad \mee & \meu & \met \\[0.2cm]
                        & \mmm & \mmt \\[0.2cm]
                        &      & \mtta \ea \right) \ls 
 \left( \bad 2 \cdot 10^{-10} & 0.4 \, (1.9) & 4.2 \cdot 10^{3}\\[0.2cm]
                              &  4.0 \cdot 10^{3} & 4.4 \cdot 10^{3} \\[0.2cm]
                              &           & 2.0 \cdot 10^{4} 
\ea \right) \rm GeV .  
\ee
A spread over 14 orders of magnitude can be seen. We state again that 
these are direct limits and the elements other than \mee  
should not be confused with their real values. 
As is evident and not surprising, the bound coming from \nnu is by far the 
best limit for an effective Majorana neutrino mass. 
One might argue that FCNC processes like $\tau \ra \mu \gamma$ 
place severe bounds on this effective masses. 
Applying the BR from \cite{mohapal} to the measured limits from 
\cite{PDG,muegammalimit} gives bounds for 
$\overline{m_{\alpha \beta}} = \sqrt{ \sum U_{\alpha m} U_{\beta m} m_m^2}$ 
of the order 1 to a few 10 GeV\@.  Without specifying to a special 
mass and mixing scheme it is rather difficult to compare 
$\mab \! \!$ with $\overline{m_{\alpha \beta}}$. Since there is no 
commonly accepted scheme around, we believe that numbers derived from  
experiments are necessary.  
Another point is that in principle one could derive the 
remaining elements of $\mab \! \!$ from the 
limits on $\meu \! \!$ and $\mee \! \!$.  
Here the same argument holds. 
One should say that if the value for \mee is fixed, all other 
elements of \mab should be in the same order of magnitude, 
therefore at most a few eV\@. Our matrix (\ref{meffresult}) is 
thus far from being physically realized.\\ 
The factor of about 3 the $(\tau \tau$) limit is worse is due to  
our condition that only the tau decay into muons is considered, which
could be skipped 
in a more general analysis including more decay channels. 
Note that with our assumption $\sigma \propto \mab \! \! ^2$ the 
bound is proportional to $\sqrt{1/\sigma \cal L}$. 
Therefore a general treatment of all possible tau decay channels would 
bring a factor of about 5.8 for $(\tau \tau$) and 2.4 for the 
channels involving only one tau. 
Furthermore, an optimistic luminosity value of 
1 fb$^{-1}$ would bring another factor of about 5.\\ 
An upgraded HERA with 
$E_p$ = 1020 GeV, $E_e$ = 33.5 GeV rises our cross sections for the small 
mass regime by
about 60 $\%$, so 
that with an integrated luminosity of 1 fb$^{-1}$ and 
in consideration of all tau channels 
our bounds could thus be lowered by a factor of 40.

\section{Conclusions}
We have studied the Majorana neutrino induced process 
$e^{\pm} p \ra \stackrel{(-)}{\nel} l^{\pm} l'^{\pm} X$ at HERA and 
deduced for the first time direct 
bounds on all effective Majorana masses other than the one 
measured in $\nnu \! \!$.  
A way to distinguish signal events from each other as well as from 
background is discussed. 
We propose a search for two like--sign muons in the final state combined with 
large missing transverse momentum. 
The cross sections are typically 12 to 13 orders of magnitude smaller 
than SM CC processes, which has to be compared 
with related rare meson decays
or $\mu^- \mu^+$ conversion on nuclei, 
which give ratios of at most $10^{-20}$.  
We improved the direct limit on 
\mmm significantly and gave for the first time direct limits on 
$\met \! \!$, \mmt and $\mtta \! \!$. However, these are direct bounds, which 
will be orders of magnitude worse than ones derived indirectly.  

\begin{center}
{\bf {\large Acknowledgments}}
\end{center}
This work has been supported in part (W.R., M.F.) by the
``Bundesministerium f\"ur Bildung, Wissenschaft, Forschung und 
Technologie'', Bonn under contract number 05HT9PEA5.  
A scholarship (W.R.) from the Graduate College 
``Erzeugung und Zerf$\ddot{\rm a}$lle von Elementarteilchen'' 
at Dortmund University is gratefully acknowledged.

\end{document}